\documentclass[preprint,aps]{revtex4-2}

\usepackage{graphicx}
\usepackage{dcolumn}
\usepackage{bm}
\usepackage{color}

\begin{document}


\title{Reply to ` ``Comment on Doubly periodic solutions of the focusing nonlinear Schrödinger equation: Recurrence, period doubling, and amplification outside the conventional modulation-instability band" '
}

\author{Matteo Conforti$^{1}$ and Nail Akhmediev$^{2}$}
\affiliation{%
$^{1}$University of Lille, CNRS, UMR 8523—PhLAM—Physique des Lasers Atomes et Mol\'ecules, Lille, France\\
$^{2}$Optical Sciences Group, Department of Theoretical Physics, Research School of Physics, The Australian National University, Canberra, ACT 2600, Australia
}
%


\begin{abstract}
We reply to the comments on our previous paper Physical Review A, Vol. 101, 023843 (2020), raised by H. Schuermann and V. Serov in 	arXiv:2204.05846.
\end{abstract}

\maketitle
The Authors of the Comment \cite{comment} raise doubts on the correctness of the solutions reported in our paper published in Phys. Rev. A \cite{Conforti2020} and the earlier paper by one of us \cite{Akh1987}.
We will show below that the criticisms of the Authors are unfounded or based on confusion. We insist that every single equation reported in \cite{Conforti2020} is correct.

Firstly, all solutions obtained in \cite{Conforti2020} can be confirmed by direct substitution into the Nonlinear Schr\"odinger Equation (NLSE) (Eq. (1) in \cite{Conforti2020}). This has been done by us as well as by the authors of many other papers on this issue published since 1987 (The most recent one is \cite{Chen2019}). In addition,
we have extensively checked the correctness of the solutions reported in \cite{Conforti2020} by numerical solution of the NLSE for a large choice of parameters. In all cases, the numerical simulations and the analytic results gave identical outcome.
Apparently, the authors of the Comment did not use multiple ways of checking their derivations in their calculations. We consider this as a deficiency of such `pure' mathematical approach.

Moreover, as physicists, we have exploited the solutions derived in \cite{Conforti2020} to model real experiments in optics and in hydrodynamics \cite{Vander2020,Vander2021}. In all cases that we have considered, the experimental results were in perfect agreement with the analytic expressions in \cite{Conforti2020}. Thus, our formulas are not only correct. They are of practical interest in describing physical effects in optics and hydrodynamics.

Below we  confute the claims of the Authors of the Comment (copied here for convenience in \emph{italics}).\\

\emph{``First, $Q(t, z)$, according to Eq.(6) in [1], is not a solution of the associated Eq.(15) in [2] (since in [1] only the final form of the solutions are presented, we refer to [2], if necessary).''}
\emph{``- $Q(t, z)$ according to Eq.(6) in [1] (Eq.(24) in [2]) is flawed.''}

The Authors of the Comment are ignoring the fact that the NLSEs in [1] and [2] have different forms. Moreover, definitions of $z$ and $t$ are also different. Therefore, 
equations cannot be simply transferred from one publication to another. The differences in definition must be taken into account.

The following two sentences in the Comment is clearly the result of the same confusion:

\emph{`` Second, consistency of Eq.(6) in [1] with Eq.(5) in [2] has not been checked neither in [1] nor in [2]. Third, solution Q(t, z) in [1] (see Fig.1) does not satisfy Eq.(5) in [2].''}

\emph{``– Consistency of Eq.(5) in system (4), (5) (in [2]) with $Q(t, z$) (in [1]) has not been checked.
Equation (6) in [1] does not satisfy Eq.(5) in [2] (or [28]) above).''}

Equations in Refs. [1] and [2] cannot be mixed. They cannot be directly transferred from Ref.[1] to Ref.[2] and vice versa because of different forms of the NLSE and different notations for $x$ and $t$ in these works. The Authors of the Comment should look carefully into this issue. The same confusion appears in the following excerpt from the Comment:

%
%

\emph{Solutions $ \delta(z)$ and $Q(t, z)$ (see Eqs.(4), (6), (19), (20) in [1]) are expressed in terms of
Jacobi elliptic functions (with $\alpha_i$ as the roots of the fourth degree polynomial, Eq.(13) in
[2]). Despite the equivalence of Jacobi and Weierstrass elliptic functions, it is not a matter
of preference to use one of both for representations of h(z) and f(t, z) (see [4], summary).
Varying parameters (e.g., W, H, D in [2]) are leading to various $\alpha_i$ and hence (in general)
to different Jacobi functions as in [1] (Eqs.(4) and (19)). Our representation of $\delta(z)$ and
$Q(t, z)$ as $h(z)$ and $f(t, z)$ according to (7) and (14) makes this discrimination unnecessary,
since (7) and (14) are valid, independently on the sign of $\Delta_z$ . Since $f(t, z)$ is triggered by $h(z)$, variable modulus is the normal case. Thus, compared with (7), (14), it seems (at least)
inexpedient to use Jacobi functions for evaluation.}

About the representation of solutions: The solutions reported by the Authors of the Comment are expressed in terms of Weierstrass elliptic functions.
This idea is not new and also comes from Ref.\cite{Akh1987} (see the end of Section 2). This alternative approach was used recently by Conte \cite{Conte}. Mathematically, the two representations are equivalent. Indeed, Weierstrass elliptic functions can be written in terms of Jacobi elliptic functions and vice versa.
The elusive `convenience' of using the Weierstrass function is that it allows to write a single expression for both type-A and type-B solutions. However, from physical point of view, this advantage is deceptive because the 
type-A and -B solutions describe qualitatively different orbits in phase space.
The two families of solutions are separated by the homoclinic trajectories known as Akhmediev breathers.
Having two different mathematical expressions for solutions describing different physical phenomena provides clarity. Unfortunately, the representation given by the Authors of the Comment does not separate the two cases and doing so would require additional formal analysis which is not given.
\\

\emph{
Solution (4) in [1] is a particular case of solution (7) ($h_0 = 0$ and three positive roots
of $R_1(h) = 0$). Additionally, due to $\delta(0) = 0$ according to (4) in [1], function value
$Q(0, 0) = Q_D = \sqrt{\alpha_1}-\sqrt{\alpha_2}-\sqrt{\alpha_3}$ (see (6) in [1]) is special compared with the range
of $f_0$ defined by constraint $R_2(f_0, z) \geq 0$ in (14). Needless to say that free (or free in a
certain domain) parameters are important for matching with experimental data; it seems
that fixed $\delta(0)$ and $Q(0, 0)$ are unnecessarily restrictive}

The Authors of the Comment suggest that their solutions are more general because they have more parameters. This is an illusion. Additional parameters can be introduced in our solutions by exploiting the well known symmetries of the NLSE which are (1) the trivial translations in space and time $z\rightarrow z +z_0$, $t\rightarrow t +t_0$, (2) the rescaling of variables, and (3) the Galilean transformation. Using them leave the NLSE unchanged. 
Adding trivial translations as suggested by the Authors of the Comment does not add any mathematical rigour into the solutions. The solutions reported in \cite{Conforti2020} 
use the minimum number of nontrivial parameters, which is three. So, fixed $\delta(0)$ and $Q(0, 0)$ \emph{ are not } really restrictions.

Talking about experiments, the solutions reported in \cite{Conforti2020} have been successfully exploited in optics \cite{Vander2020, Vander2021}. Better than this, they permitted to discover a novel form of modulation instability beyond the conventional (linear) instability band \cite{Vander2021}. The latter is a significant extension of the well known  Benjamin-Feir instability. 
\\

\emph{Finally, $f(t, z)$ is different from $Q(t, z)$ because the modulus $k_q$ of $Q(t, z)$ does not depend on $z$. $Q(t, z)$  is the solution of Eq.(15) in [2] (with disregards to different notations of
the variables), which is equation (5) above. The coefficients in both equations are dependent
on $z$ and (on $t$ in [2]). Thus it is not clear why the $z$-dependence drops out from the
modulus of Eq.(6) in [1] (Eq.(24) in [2]). Moreover, the validity of solution (6) in [1] is not
justified: If we assume that $Q(t, z$) according to (6) in [1] is a solution of the corresponding
equation (15) in [2] then the modulus $k_q$ in (6) is not correct. If we assume that $Q(t, z)$ with modulus $k_q$ is correct then $Q(t, z)$ is not a solution of equation (15) in [2].}

Again, the Authors of the Comment are confusing the matter trying to mix  equations in [1] and [2]. We have to repeat that the forms of the NLSE in Refs.[1] and [2] are different and the definition of variables $x$ and $t$ are different. The notations in [1] are common in optics while the notations in [2] are common in mathematical literature.
\\

\emph{IV. ON THE ADEQUACY OF ANSATZ (2)\\
As mentioned in the Introduction, consistency of $Q(t, z)$ (according to (6) in [1]) with
Eq.(5) in [2] has not been checked. Thus it leads to the question whether f(t, z) according
to (14) is consistent with Eq.(3a).}

The title of this section in the Comments is bizarre. It assumes that the basic conjecture used in [1] (and its equivalent in [2]) is incorrect. In reality,
the fact that the class of solutions derived in \cite{Conforti2020} and \cite{Akh1987} satisfy the ansatz
\begin{equation}\label{ansatz}
\psi(t,z)=[Q(t,z)+i\delta(z)]e^{i\phi(z)}.
\end{equation}
can easily be confirmed by simple analysis of these solutions. Namely, one-soliton solutions, Akhmediev breathers, Kutznetzov-Ma solitons, variety of cnodial waves and the lowest order rogue wave all belong to this class \cite{Akh2009}. This is the matter of simple algebra. As explained in \cite{Akh1987}, Eq.(\ref{ansatz}) means that at any fixed $z$ (or $t$ in [2]), the real and imaginary parts of these solutions are linearly related. Then, it is unclear what the Authors of the Comment mean by the `adequacy of the ansatz'. 
Rejecting the existence of the ansatz is equivalent to claiming that the above class of solutions do not exist.

%

Our procedure is straightforward and common in physics: we made a conjecture [Eq. (\ref{ansatz})], we found solutions and we checked them extensively by different means (direct substitution, numerical simulations of the NLSE, and confirmed the mathematical results in experiments). \\

\emph{The representation of $Q(t, z)$ in terms of Jacobi functions is not effective for numerical
evaluation.}

This claim is simply wrong. Firstly, the Jacobi elliptic functions today are as common as the trigonometric functions. They are well familiar to most students specialised in physics or mathematics. Such presentation demonstrates clearly the qualitative difference between the type-A and the type-B solutions imposed by the physics of the problem. Secondly, software packages for calculating elliptic Jacobi functions are common and widely available. This cannot be said about the Weierstrass function. For example, MATLAB software implements elliptic Jacobi functions but not the Weierstrass function. Thirdly, we addressed the solutions reported in \cite{Conforti2020} to physicists and experts in nonlinear dynamics. Thus, we limited ourselves to the bounded, smooth and non-singular solutions and illustrated them graphically. 
What is presented in the Comments are  manipulations with equations that can easily go wrong without additional confirmation using either numerical calculations or a simple technique such as direct substitution of solutions into the original equation. 

To conclude, we have confuted point-by-point the criticisms of the Authors of the Comment. The correctness of the doubly periodic solutions reported in \cite{Conforti2020} is easy to check using independent techniques such as a simple substitution into the NLSE. This is exactly what has been done. While we are sure about the correctness of our solutions, we have doubts on the expressions reported in the Comment. Finding the actual errors in their derivations is a task that goes well beyond the scope of this Reply.

In our opinion the Comment by Sch\"urmann and Serov contains several false statements. Moreover, it does not report any novelty: the solutions in terms of Weierstrass functions have been first suggested in \cite{Akh1987} and have been recently implemented in more details by Conte \cite{Conte} (ref. [10] of the Comment). Remarkably, the new results (as claimed) in \cite{Conte} are singular solutions of the NLSE that have been omitted completely in \cite{Conforti2020} and \cite{Akh1987} as they are unphysical. We should also note that although the major part of derivations in \cite{Conte} have been copied from \cite{Conforti2020} and \cite{Akh1987}, no any errors were detected. This is in striking contradiction with the claims of the Authors of the Comment.


\begin{thebibliography}{10}


\bibitem{Conforti2020}
M. Conforti, A. Mussot, A. Kudlinski, S. Trillo, and N. Akhmediev, "Doubly periodic solutions of the focusing nonlinear Schrödinger equation: Recurrence, period doubling, and amplification outside the conventional modulation-instability band," Phys. Rev. A 101, 023843 (2020).

\bibitem{Akh1987}
N. N. Akhmediev, V. M. Eleonskii, and N. E. Kulagin, "Exact first-order solutions of the nonlinear Schrodinger equation," Theor. Math. Phys. 72, 809–818 (1987).

\bibitem{Vander2021}
G. Vanderhaegen, C. Naveau, P. Szriftgiser, A. Kudlinski, M. Conforti, A. Mussot, M. Onorato, S. Trillo, A. Chabchoub, and N. Akhmediev, "“Extraordinary” modulation instability in optics and hydrodynamics," Proc. Natl. Acad. Sci. 118, e2019348118 (2021).

\bibitem{Vander2020}
G. Vanderhaegen, P. Szriftgiser, C. Naveau, A. Kudlinski, M. Conforti, S. Trillo, N. Akhmediev, and A. Mussot, "Observation of doubly periodic solutions of the nonlinear Schrödinger equation in optical fibers," Opt. Lett. 45, 3757 (2020).

\bibitem{Chen2019}
J. Chen, D. E. Pelinovsky, and R. E. White, "Rogue waves on the double-periodic background in the focusing nonlinear Schrödinger equation," Phys. Rev. E 100, 052219 (2019).

\bibitem{Akh2009}
N. Akhmediev, J. M. Soto-Crespo, and A. Ankiewicz, "Extreme waves that appear from nowhere: On the nature of rogue waves," Phys. Lett. A 373, 2137–2145 (2009).

\bibitem{Conte} R. Conte, Theor. Math. Phys., {\bf 201}, 1366 (2021). 

\bibitem{comment}
H. W. Schuermann and V. Serov, "Comment on "Doubly periodic solutions of the focusing nonlinear Schrödinger equation: Recurrence, period doubling, and amplification outside the conventional modulation-instability band"," 	arXiv:2204.05846 (2022).


\end{thebibliography}
\end{document}